\documentclass[a4paper,11pt]{article}

\usepackage{pos}
\usepackage{wrapfig}
\usepackage{lipsum} 
\usepackage{amsmath}
\usepackage{lineno}

\title{A model independent parametrization of the optical properties of the refrozen IceCube drill holes}

\ShortTitle{A model independent parametrization of the optical properties of the refrozen IceCube drill holes}

\author{The IceCube Collaboration \\{\normalsize \normalfont(a complete list of authors can be found at the end of the proceedings)}\\}

\emailAdd{philipp.eller@tum.de}
\emailAdd{martin.rongen@fau.de}

\abstract{

The IceCube Neutrino Observatory deployed 5160 digital optical modules (DOMs) in a cubic kilometer of deep, glacial ice below the geographic South Pole, recording the Cherenkov light of passing charged particles. While the optical properties of the undisturbed ice are nowadays well understood, the properties of the refrozen drill holes still pose a challenge. From camera observations, we expect a central, strongly scattering column shadowing a part of the DOMs' sensitive area. In MC simulation, this effect is commonly modeled as a modification to the DOMs' angular acceptance curve, reducing the forward sensitivity of the DOMs. The associated uncertainty is a dominant detector systematic for neutrino oscillation studies as well as high-energy cascade reconstructions. Over the years, several measurements and fits of the drill holes' optical properties and of the angular acceptance curve have been proposed, some of which are in tension. Here, we present a principle component analysis, which allows us to interpolate between all suggested scenarios, and thus provide a complete systematic variation within a unified framework at analysis level.

\vspace{4mm}
{\bfseries Corresponding authors:}
Philipp Eller$^{1*}$, Martin Rongen$^{2}$\\
{$^{1}$ \itshape Technical University of Munich, TUM School of Natural Sciences, Physics Department, 85747 Garching, Germany}\\
{$^{2}$ \itshape Erlangen Centre for Astroparticle Physics, Friedrich-Alexander Universität Erlangen-Nürnberg}\\[4mm]
$^*$ Presenter

\ConferenceLogo{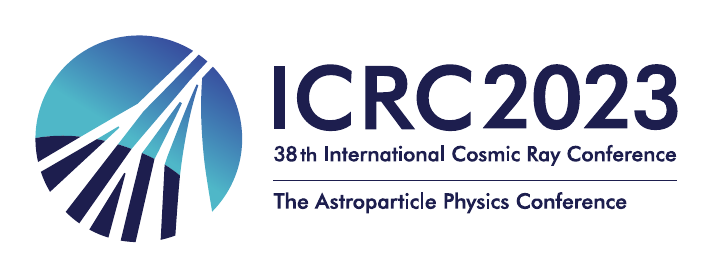}

\FullConference{The 38th International Cosmic Ray Conference (ICRC2023)\\ 26 July -- 3 August, 2023\\ Nagoya, Japan}
}

\begin{document}

\maketitle

\section{Introduction}\label{sec1}

The IceCube Neutrino Observatory~\citep{DetectorPaper} is located in the deep glacial ice below the geographic South Pole. It is comprised of 5160 \textit{Digital Optical Modules (DOMs)}\, each incorporating a 10-inch, downward facing photomultiplier tube to detect Cherenkov light produced by relativistic charged particles traversing the detector. The DOMs were deployed along so called \textit{strings} in 86 drill holes of $\sim$60\,cm diameter at depths spanning between 1450\,m and 2450\,m. While the dominant part of the propagation of Cherenkov photons from their emission to an eventual detection happens in the exceptionally clear, bulk glacial ice, each photon detected by a DOM also had to propagate through the refrozen water in a drill hole, the so called \textit{hole ice}. This hole ice is modeled as a variation of the optical acceptance of DOMs as a function of the photon incident zenith angle.

While the calibration and modeling of the optical properties of the bulk glacial ice has continuously improved over the years \cite{Flasher2013, ICRCanisotropy, tc-2022-174}, the optical properties of the hole ice are less well understood.
Years ago the uncertainties in most IceCube analyses were dominated by statistics, today for example the DeepCore oscillation analyses \cite{Aartsen_2018, IceCube:2019dqi, IceCube:2023ewe} are limited by uncertainties of detector effects such as bulk ice properties and the overall optical efficiency of the modules, but also especially the angular acceptance of modules that is discussed here.
Since optical effects have to be modeled in simulation by altering the photon propagation and acceptance, incorporating such effects in analyses as nuisance parameters is computationally costly. The usual approach is to either build interpolating functions between discrete simulation sets (e.g. \cite{Fischer:2023dbo}) or using random sampling of simulation configurations (e.g. \cite{IceCube:2019lxi}).

We lay out three properties a good model should fulfill:
\begin{enumerate}
    \item Provide enough flexibility to cover a large enough range of scenarios to be able to describe the data
    \item Use as few parameters as possible to reduce computational burden on simulation and inference
    \item Define a specified range of parameter values to inform the simulation to cover a useful space of possibilities
\end{enumerate}

These points are addressed in this work, and documented in the following sections: Section~\ref{sec:old} summarizes several measurements and fits of the drill holes' optical properties and of the angular acceptance curve that have been proposed in the past. In Sec.~\ref{sec3} we introduce our new two-parameter model that allows us to interpolate between all suggested scenarios, and thus provide a complete systematic variation within a unified framework at analysis level. The parameters of this model are fitted to LED calibration data in Sec.~\ref{sec:flasher}. Finally a comparative overview over input models, the results from the LED fit as well as results from nuisance parameter fits in physics analyses using the described model is provided.

\section{Previous Models}\label{sec:old}

\begin{figure}
    \centering
    \includegraphics[width=0.6\textwidth]{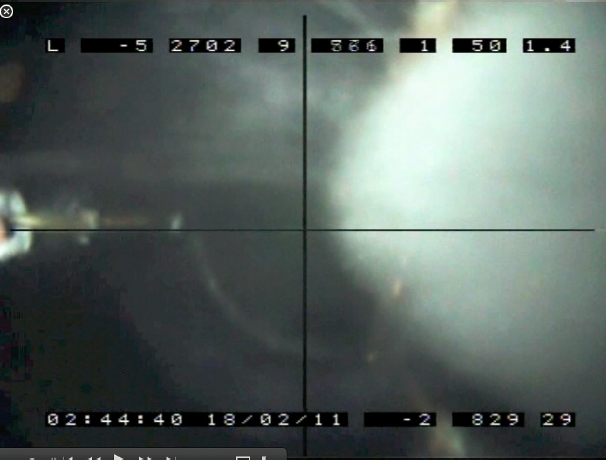}
    \caption{Sweden Camera image of the lower camera looking straight up at the upper camera, with the hole ice fully developed in the right half of the image. \cite{Hulth}}
    \label{fig:SwedenCamera}
\end{figure}

Direct visual evidence for the existence and properties of the refrozen drill columns has been provided by a pair of cameras situated at the bottom of a string~\cite{DetectorPaper}. As seen in  Figure \ref{fig:SwedenCamera}, the outer regions of the drill hole are exceptionally clear, but the inner $\sim$16\,cm appear diffuse white, as would be expected from strong scattering on air bubbles. This concurs with a model where the water filled holes freeze cylindrically inwards with impurities, including air bubbles, being continuously pushed ahead of the freezing boundary until they precipitate out in high concentration in the center of the hole. This feature is commonly denoted the \textit{bubble column}.

In simulation the bubble column can be included directly by physically modeling the photocathode extend as well as an extended bubble column with assumed diameter, scattering properties and position with respect to each DOM. This very precise approach has recently been pursued successfully in \cite{MartinPHD} and  \cite{DimaICRC21}, but is computationally expensive. 
Thus, the effect induced by the bubble column is instead commonly modeled via an angular acceptance curve. In this approximation the landing position of a photon on the surface of a DOM is disregarded when evaluating the detection probability in simulation and instead the efficiency is only based on the cosine of the zenith angle of the photon direction at the time of impact on the module ($\cos\eta)$.

\begin{figure}[h]
    \centering
    \includegraphics[width=0.75\textwidth]{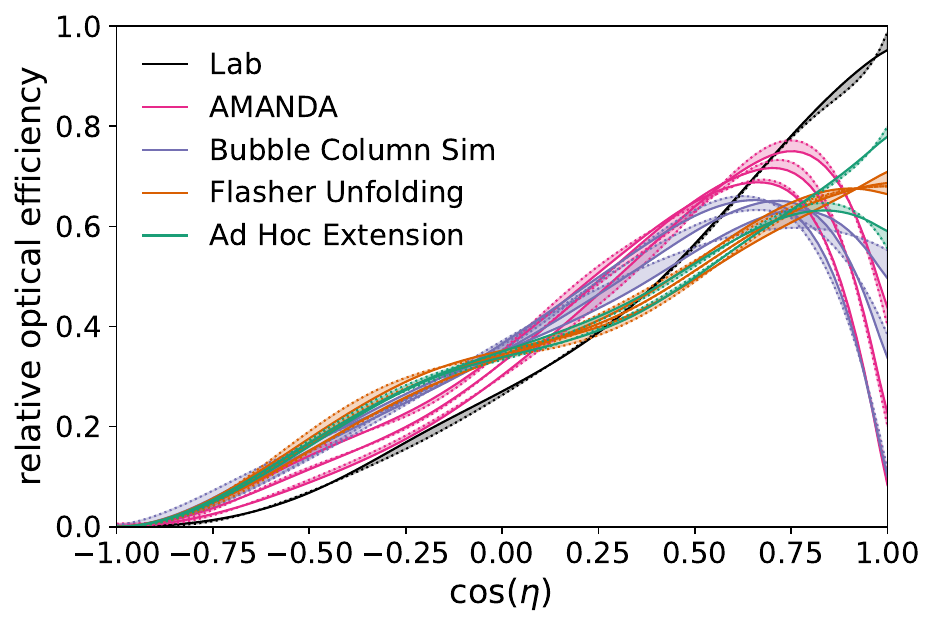}
    \caption{DOM angular acceptance curves from previous modeling attempts, used as input for the Unified Model described in this work. The acceptances are shown as functions of the PMT incident angle $\eta$, where $\cos{\eta}=1$ means photons coming head-on to the photocathode area. \textit{Lab} denotes the DOM characteristics without hole ice. \textit{AMANDA} refers to measurements performed by a laser system in the IceCube predecessor AMANDA. \textit{Flasher Unfolding} refers to a constrained unfolding using IceCube LED calibration data. \textit{Ad Hoc Extension} is a parametric extension of this model used in previous neutrino oscillation fit. And \textit{Bubble column Simulation} is a set of curves derived from a first principle simulation of photon propagation through allowed hole ice configurations.
    In each case the dotted line shows the original input model, while the solid line shows the closest representation of this input within the two parameters of the Unified model.}
    \label{fig:oldmodels}
\end{figure}

Approximating a DOM to be homogeneously sensitive on its entire lower hemisphere, its bare relative detection efficiency ($y$) is then given as $y_{\mbox{ideal}}=0.5\cdot(1+\cos\eta)$. The angular acceptance curve resulting from a precise knowledge of the DOM's hardware properties
~\cite{Flasher2013, MartinPHD} is denoted \textit{Lab} in Figure \ref{fig:oldmodels}. Incorporating the effect of the hole ice into the angular acceptance generally decreases the forward acceptance at $\cos\eta=1$, as the hole ice effectively shadows off a part of the photocathode.

The earliest, but still commonly used models, were derived for the IceCube predecessor AMANDA~\cite{Hfamily, MartinPHD}. Here, lacking the later visual camera evidence, it was assumed that the entire $\sim$60\,cm drill hole would have degraded optical properties with geometric
scattering length in the range between 50\,cm and 100\,cm.

Within IceCube the optical properties are most commonly deduced from fits to LED calibration data, which have little direct sensitivity to the forward acceptance. For this reason, previous fits~\cite{MartinPHD} used the following functional form for the angular acceptance 
\begin{equation}
    y=0.34 (1+1.5\cos\eta-\cos^3\eta/2)+p \cdot \cos\eta (\cos^2\eta-1)^3.
\end{equation}
It is constrained to a reasonable value for the forward acceptance, roughly shading of 30\% of the photocathode area, and the free parameter $p$ only changes the shape for intermediate values of $\cos\eta$.

Yet, as the forward acceptance is of primary importance as a systematic uncertainty for analyses, this constraint was lifted in an ad-hoc parametric extension, which allows for an arbitrary forward acceptance through a second parameter~\cite{PhysRevLett.120.071801}: 
\begin{equation}
    y=0.34 (1+1.5\cos\eta-\cos^3\eta/2)+p \cdot \cos\eta (\cos^2\eta-1)^3 + p2 \cdot \exp(10(\cos\eta-1.2)).
\end{equation}

While the previously mentioned direct simulation of photon propagation through a bubble column is conceptually different from the angular acceptance parametrization, the hole ice properties (size and scattering length) deduced there can also be translated into angular acceptance curves assuming the bubble column to be centered on all DOMs. These curves, with assumed diameters ranging between 18\,cm and 54\,cm and effective scattering lengths ranging between 14\,cm and 125\,cm respectively are denoted as \textit{Bubble Column Simulation} in the following. The allowed parameters were derived in \cite{MartinPHD} and have since been superseded by \cite{DimaICRC21}. As the update primarily affected the per-DOM positions of the bubble column, but not the size and scattering length, the curves are still assumed to represent the best knowledge at present. 

\section{Unified Model}\label{sec3}

\begin{figure}[h]
    \centering
    \includegraphics[width=0.85\textwidth]{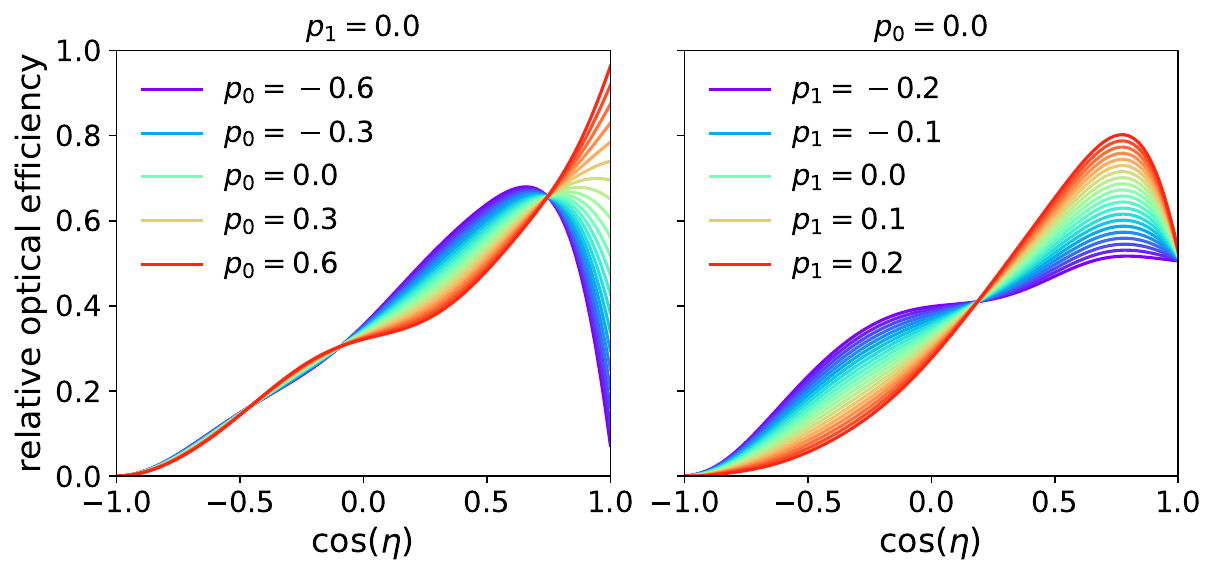}
    \caption{Example angular acceptance curves, resulting from variations of the $p_0$ and $p_1$ parameters of the Unified model. $p_0$ primarily suppresses the forward-acceptance and thus models the fraction of shadowed photocathode area, while $p_1$ primarily modifies the slope and thus isotropy of detection.}
    \label{fig:PCAcurves}
\end{figure}

The aim of the unified model is to define a parameterization with as few parameters as possible that can approximate all existing angular acceptance curves as discussed in the previous section and shown in Fig.~\ref{fig:PCAcurves}, as well as interpolate between those and extrapolate outside. Note that the ad-hoc model presented in the previous section is also based on just two parameters, but it does not offer enough flexibility to approximate other models and variations of its acceptance curves are constrained to a small region.

We begin by approximating the existing curves using an interpolating B-spline of order $k=3$ (cubic).
Seven support points for the spline are defined at:
\begin{equation}
    x=\cos{\eta} = (-1,-0.5,-0.2,0.35,0.65,0.95,1.05),
\end{equation}
where the last point extends outside the physically allowed region. These values have been heuristically determined to work well.
The corresponding points in $y$ (= relative optical efficiency) are free parameters except for the points at $x=-1$ where the value is fixed to $y=0$, since photons arriving from behind the PMT can not be detected.
Furthermore, derivatives are fixed to zero at either end of the spline.
This means that with with the six parameters (degrees of freedom) constituting $y$ we define B-splines that make up angular acceptance curves.
In a fit to the original twelve curves discussed in the previous section, the best values for $y$ are determined by minimizing the mean-squared error between the original curves and their B-spline approximations.
This means that all models can be described by a matrix $\mathbf{S}$ of $(12 \times 6)$ values.

In order to facilitate the usage of the new model in physics analyses, we want to reduce the dimensionality as much as possible, to end up with a minimal number of nuisance parameters to vary in simulation and inference.
We apply a singular value decomposition (SVD) to the matrix $\mathbf{S}$ to express it in the form $\mathbf{U} \mathbf{\Sigma} \mathbf{V}^*$. This can then be interpreted as a set of six principal components $p$ per calibration curve collected in the matrix $\mathbf{U} \mathbf{\Sigma}$ and six principal directions arranged in matrix $\textbf{V}$.

Reducing the number of principal components and accordingly the components of the direction vectors is known as a principal component analysis (PCA) allowing to reduce dimensionality.
It turns out that by only using the first two components, henceforth referred to as $p_0$ and $p_1$, is sufficient to reconstruct all input curves to a precision of $<5\%$.
The fitted principal directions are given below:
\begingroup
\begin{equation}
\mathbf{V}^* = 
\begin{pmatrix}
-0.0054533 &  0.0165525 & -0.136688 & -0.0782252 & 0.5139002 &  0.8430896\\
-0.4508718 & -0.5179942 &  0.2407660 &  0.6061325 & 0.3092565 & -0.0859777
\end{pmatrix}
\end{equation}
\endgroup

\noindent and the original points $y$ needed for the B-spline can then be reconstructed as:
\begin{equation}    
y - \langle y \rangle = (p_0, p_1) \cdot \mathbf{V}^*
\end{equation}

\noindent where the mean that was subtracted in the decomposition is equal to:
\begin{equation}    
\langle y \rangle  = (0.14623077, 0.26576604, 0.4832101 , 0.63038745, 0.57494938, 0.48991044).
\end{equation}

Having a vector $y$ at hand given a choice of parameters $(p_0, p_1)$ then allows to construct the B-spline describing the angular acceptance curve.
This curve is then in further processing steps clipped to be greater or equal to zero everywhere, and then normalized to equal area under the curve such that all curves reflect the same total efficiency.

In summary, the outlined procedure defines a two-parameter unified model that can generate an angular acceptance curve for a choice of input parameters $(p_0, p_1)$, and is able to approximate all existing calibration curves as discussed in the previous section to an accuracy of $<5\%$.

Figure \ref{fig:PCAcurves} shows some example curves for varying inputs of $p_0$ and $p_1$, respectively. It can be noted that the first component $p_0$ results in a modulation of only the head-on region of the PMT ($\cos{\eta} \gtrapprox 0$), while component $p_1$ affects mostly the shape around the waist of the PMT.
Figure \ref{fig:oldmodels} shows the fidelity of the 2-parameter approximation of the unified model compared to the initial calibration curves used as inputs.
The maximum range of the parameters that produces sensible curves is roughly $-2 < p_0 < 1$ and $-0.2 < p_1 < 0.2$.

\section{Flasher fit}\label{sec:flasher}

While the parametrization has been derived from plausible models resulting from fits to muon and LED calibration data, its parameters can also be fitted directly. This in particular allows to  constrain a sensible region for large-scale simulations to be conducted. 

\begin{figure}
    \centering
    \includegraphics[width=0.9\textwidth]{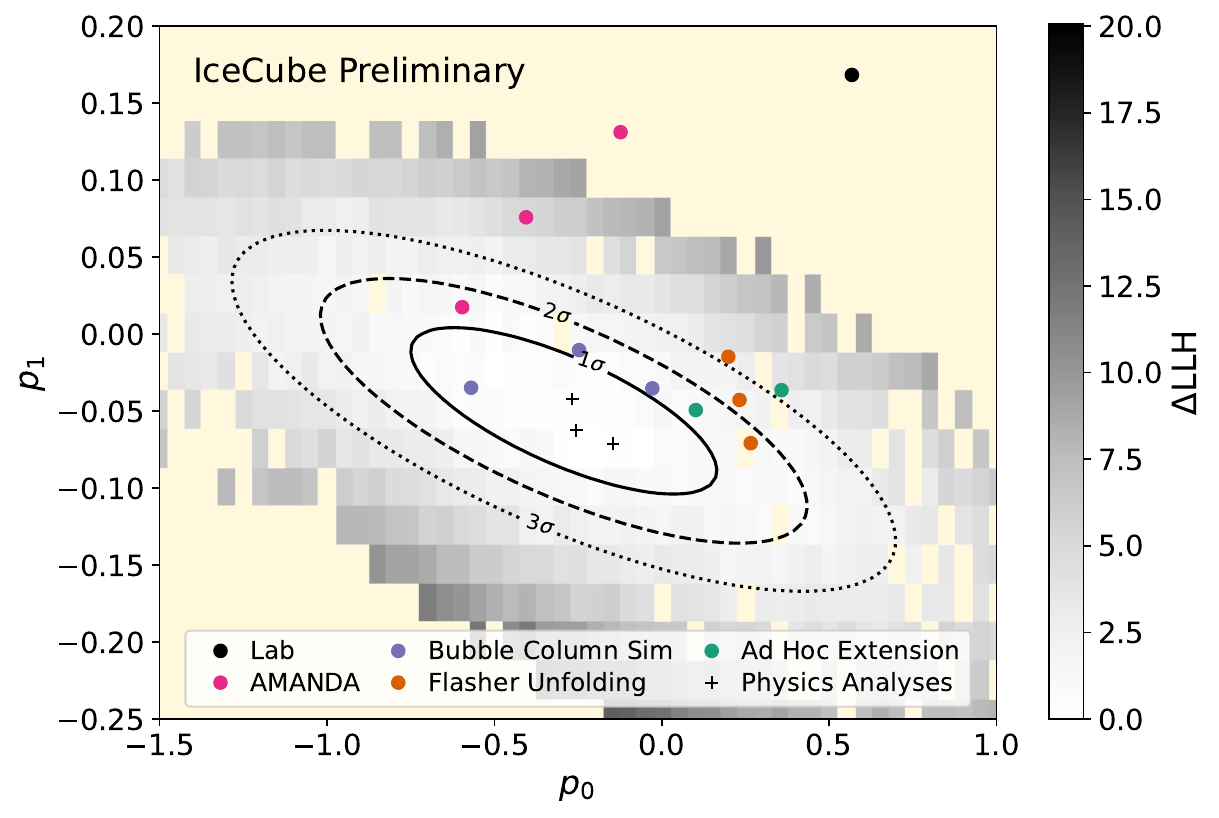}
    \caption{$p_0$-$p_1$-parameter landscape of the Unified Model. Colored circles denote the parameter values of the input models used to build the parametrization. The intensity coded grid cells show the delta likelihood value of a flasher fit for the given simulated hole ice realization. The fit was performed using 120 light emitting DOMs surrounding DeepCore at depths greater then 2100\,m. The contours denote the statistical error only and are dominated by the simulation statistics. The black crosses denote the preferred values as resulting from nuisance parameter fits in several low- and high-energy physics analyses.}
    \label{fig:landscape}
\end{figure}

To this end a fit to LED calibration data, selecting data from DOMs on six standard IceCube strings surrounding DeepCore and at depth exceeding 2100\,m, each flashing their 12 LEDs~\citep{Aartsen_2017} in sequence, has been performed. The fit follows the methodology as described in \cite{Flasher2013, MartinPHD}, comparing the arrival time distributions in data to photon propagation simulation~\citep{photonpropconference}, with only the hole ice parameters being varied in simulation. All other aspects and parameter values of the ice optical modeling are taken from the so-called Spice3.2 ice model~\citep{MartinPHD}.

Figure \ref{fig:landscape} shows the resulting likelihood space, with each grid cell being intensity coded according to the distance of the likelihood value of one simulated hole ice realization from the best-fit realization. The employed likelihood (see \cite{chirkin2013likelihood}) accounts for the vastly smaller photon statistics in simulation compared to the experimental data. This induces fluctuations of the likelihood values compared to the expected paraboloid. The statistics-only uncertainty contours as shown account for this fluctuation by fitting a polynomial. As the likelihood does not conform to Wilk's Theorem the $\Delta LLH$ values for a given coverage have been calculated from the residuals around this polynomial. The contour sizes primarily reflect the employed simulation statistics, but are representative of the sensitivity of the analysis as a whole. The impact of other systematic detector uncertainties (such as the bulk ice modeling) on the preferred hole ice parameters has not been evaluated in this context, as it would require equivalent fits and thus vastly more simulation.

\section{Results \& Discussion}\label{sec:results}

Figure~\ref{fig:landscape} shows the values (black crosses) of the maximum likelihood estimators for $p_0$ and $p_1$ in three recent IceCube physics analyses that use our model. Interestingly, these values coincide with the minimum of the fit to flasher data, drawing a consistent picture.
The fit to flasher data also can make some statements on previously used models, and strongly disfavors the scenario of absence of hole ice, i.e. no altered optical properties in the refrozen drill holes corresponding to the "Lab" curve.
Furthermore, the AMANDA models do not seem to be able to correctly describe the IceCube angular acceptance. The previous attempts using simple parametrizations ("Flasher unfolding" and its "Ad Hoc Extension") are also disfavored, albeit at a milder level. The only compatible scenarios seem to be the those derived from the direct simulation of photon propagation in the bubble column termed "Bubble column simulation" above.

Another important aspect is the range of useful parameters, that can also be read off from Fig.~\ref{fig:landscape}. For a simple rectangular box, we propose the range of $-1 < p_0 < 0.4$ and $ -0.14 < p_1 < 0.04$ roughly covering the $2\sigma$ contour. Since the likelihood exhibits a negative correlation coefficient of $\rho=-0.72$ a range of simulations more closely following the contour shape is recommended.

The IceCube Upgrade~\cite{Upgrade} will, among many more optical sensors, also deploy several new calibration hardware. The Precision optical Calibration Module (POCAM) \cite{henningsen2020self,Henningsen_2023}, for example, will allow to collect new calibration data of unprecedented quality. This data is expected to allow us to better understand our detector, including the hole ice and associated angular acceptance.

\bibliographystyle{ICRC}
\bibliography{references}

%

\clearpage

\section*{Full Author List: IceCube Collaboration}

\scriptsize
\noindent
R. Abbasi$^{17}$,
M. Ackermann$^{63}$,
J. Adams$^{18}$,
S. K. Agarwalla$^{40,\: 64}$,
J. A. Aguilar$^{12}$,
M. Ahlers$^{22}$,
J.M. Alameddine$^{23}$,
N. M. Amin$^{44}$,
K. Andeen$^{42}$,
G. Anton$^{26}$
% ...
\\
\\
% ...
$^{12}$ Universit{\'e} Libre de Bruxelles, Science Faculty CP230, B-1050 Brussels, Belgium \\
% ...
$^{17}$ Department of Physics, Loyola University Chicago, Chicago, IL 60660, USA \\
$^{18}$ Dept. of Physics and Astronomy, University of Canterbury, Private Bag 4800, Christchurch, New Zealand \\
% ...
$^{22}$ Niels Bohr Institute, University of Copenhagen, DK-2100 Copenhagen, Denmark \\
$^{23}$ Dept. of Physics, TU Dortmund University, D-44221 Dortmund, Germany \\
% ...
$^{26}$ Erlangen Centre for Astroparticle Physics, Friedrich-Alexander-Universit{\"a}t Erlangen-N{\"u}rnberg, D-91058 Erlangen, Germany \\
% ...
$^{40}$ Dept. of Physics and Wisconsin IceCube Particle Astrophysics Center, University of Wisconsin{\textendash}Madison, Madison, WI 53706, USA \\
% ...
$^{42}$ Department of Physics, Marquette University, Milwaukee, WI, 53201, USA \\
% ...
$^{44}$ Bartol Research Institute and Dept. of Physics and Astronomy, University of Delaware, Newark, DE 19716, USA \\
% ...
$^{63}$ Deutsches Elektronen-Synchrotron DESY, Platanenallee 6, 15738 Zeuthen, Germany  \\
$^{64}$ Institute of Physics, Sachivalaya Marg, Sainik School Post, Bhubaneswar 751005, India \\
% ...

\subsection*{Acknowledgements}

\noindent
The authors gratefully acknowledge the support from the following agencies and institutions:
USA {\textendash} U.S. National Science Foundation-Office of Polar Programs,
U.S. National Science Foundation-Physics Division,
U.S. National Science Foundation-EPSCoR,
Wisconsin Alumni Research Foundation,
Center for High Throughput Computing (CHTC) at the University of Wisconsin{\textendash}Madison,
Open Science Grid (OSG),
Advanced Cyberinfrastructure Coordination Ecosystem: Services {\&} Support (ACCESS),
Frontera computing project at the Texas Advanced Computing Center,
U.S. Department of Energy-National Energy Research Scientific Computing Center,
Particle astrophysics research computing center at the University of Maryland,
Institute for Cyber-Enabled Research at Michigan State University,
and Astroparticle physics computational facility at Marquette University;
Belgium {\textendash} Funds for Scientific Research (FRS-FNRS and FWO),
FWO Odysseus and Big Science programmes,
and Belgian Federal Science Policy Office (Belspo);
Germany {\textendash} Bundesministerium f{\"u}r Bildung und Forschung (BMBF),
Deutsche Forschungsgemeinschaft (DFG),
Helmholtz Alliance for Astroparticle Physics (HAP),
Initiative and Networking Fund of the Helmholtz Association,
Deutsches Elektronen Synchrotron (DESY),
and High Performance Computing cluster of the RWTH Aachen;
Sweden {\textendash} Swedish Research Council,
Swedish Polar Research Secretariat,
Swedish National Infrastructure for Computing (SNIC),
and Knut and Alice Wallenberg Foundation;
European Union {\textendash} EGI Advanced Computing for research;
Australia {\textendash} Australian Research Council;
Canada {\textendash} Natural Sciences and Engineering Research Council of Canada,
Calcul Qu{\'e}bec, Compute Ontario, Canada Foundation for Innovation, WestGrid, and Compute Canada;
Denmark {\textendash} Villum Fonden, Carlsberg Foundation, and European Commission;
New Zealand {\textendash} Marsden Fund;
Japan {\textendash} Japan Society for Promotion of Science (JSPS)
and Institute for Global Prominent Research (IGPR) of Chiba University;
Korea {\textendash} National Research Foundation of Korea (NRF);
Switzerland {\textendash} Swiss National Science Foundation (SNSF);
United Kingdom {\textendash} Department of Physics, University of Oxford.

\end{document}